\documentclass[letterpaper,11pt]{article}
\usepackage[left=1 in, right=1in, top=1in, bottom=1in]{geometry}
\usepackage[utf8]{inputenc} 
\usepackage{graphicx}
\usepackage{subcaption}
\usepackage{float}
\usepackage{setspace}
\usepackage[authoryear]{natbib}
\usepackage{wrapfig}
\usepackage{xcolor}
\usepackage{dcolumn}
\usepackage[english]{babel}
\usepackage{authblk}
\usepackage{xr-hyper}
\usepackage{pdfpages}

\usepackage{lineno}
\usepackage{url}

\begin{document}
\title{Speciation by local adaptation and isolation by distance in extended environments}

\author[1]{Lara D. Hissa} 
%\author[*]{~} 
\author[1,*]{Marcus A. M. de Aguiar}
%\author[*]{~}
\author[2]{Flavia M. D. Marquitti}
%\author[*]{~}

%\footnote{Corresponding author. E-mail: aguiar@ifi.unicamp.br}}

\affil[1]{Instituto de F\'isica `Gleb Wataghin', Universidade Estadual de Campinas, Unicamp 13083-970, Campinas, SP, Brazil}
%\affil[*]{~}
%\affil[*]{~}
\affil[2]{Instituto de Biologia, Universidade Estadual de Campinas, Unicamp 13083-970, Campinas, SP, Brazil}
%\affil[*]{~}
%\affil[*]{~}
\affil[*]{Corresponding author. E-mail: aguiar@ifi.unicamp.br}
%\affil[*]{Corresponding author. E-mail:}
%\affil{Instituto de F\'isica `Gleb Wataghin', Universidade Estadual de Campinas, Unicamp 13083-970, Campinas, SP, Brazil}

\maketitle

\begin{abstract}
Speciation is often associated with geographical barriers that limit gene flow. However, species can also emerge in parapatry, even in homogeneous environments, through spatially restricted mating and limited dispersal. When the environment is not homogeneous, natural selection contributes to differentiation by local adaptation and tends to facilitate speciation. To explore how isolation by distance and adaptation combine to determine species diversity, we propose a model regulated by these two components. The former is implemented via mating restrictions on spatial proximity and genetic similarity, whereas the latter is realized by an ecological phenotype subjected to adaptation by natural selection. We consider a scenario where the environment has two distinct optima, and compare the resulting diversity patterns and phenotypic distributions with those of a homogeneous environment, with a single ecological optimum. We show that the interplay between selection and isolation by distance affects not only species formation but also phenotypic distributions and the timing of diversification. Simulating individuals with either restrictive or permissive mating regimes, combined with strong or weak selection, we show that: (i) environmental selection can accelerate diversification but is not always necessary for reproductive isolation; (ii) bimodal phenotypic patterns can arise through different evolutionary pathways and are therefore not necessarily signatures of ecological speciation; and (iii) when selection is weak and mating is restrictive, parapatric speciation begins before pronounced ecological differentiation, while the phenotypic peaks can oscillate and fail to reach a stationary state.\\

\end{abstract}


\newpage

\setstretch{1.5}

% %%%%%%%%%%%%%%%%%%%%%%%%%%%%%%%%%%%%%%%%%%%%%%%%%%%%
% %%%%%%%%%%%%%%%%%%%%%%%%%%%%%%%%%%%%%%%%%%%%%%%%%%%%

\section*{Introduction}
Speciation requires the evolution of barriers to gene flow, yet the mechanisms responsible for generating those barriers remain a central question in evolutionary biology \citep{coyne2004speciation, mayr2013animal, nosil2012ecological}. Two broad classes of mechanisms have been particularly influential in theoretical studies. The first emphasizes spatial processes, in which limited dispersal and local mating generate genetic structure and  may allow reproductive isolation to emerge even in geographically continuous populations \citep{white1968models, gavrilets1998rapid, hoelzer2008, de2009global, martins2013}. The second emphasizes ecological divergence, whereby natural selection across heterogeneous environments promotes local adaptation and reduces effective gene flow between populations \citep{schluter2009evidence, nosil2012ecological, rundle2005ecological}. Although both mechanisms can promote diversification, determining whether reproductive isolation is primarily driven by spatial processes, environmental selection, or their interaction remains challenging \citep{smadja2011framework, fitzpatrick2008if}.

A persistent difficulty is that many models of ecological speciation couple ecological adaptation to the evolution of reproductive isolation because ecological traits directly influence mate choice, habitat preference, competition, or reinforcement \citep{servedio2011magic, kirkpatrick2002speciation, seehausen2004hybridization, nosil2012ecological, Dieckmann_2003}. Adaptation and reproductive isolation may therefore evolve simultaneously. Under these conditions, phenotypic divergence across environments, (multi-)bimodal trait distributions, and correlations between ecological and genetic differentiation are frequently interpreted as signatures of ecological speciation \citep{schluter2009evidence, nosil2009ecological, sobel2010biology}. However, these patterns do not uniquely identify the underlying mechanism, because the contemporary association between ecological differentiation and reproductive isolation may reflect a complex evolutionary history and does not, by itself, reveal the order in which these processes evolved \citep{czekanski2019ecology, bernatchez2010origin}.

This issue is particularly relevant in spatially continuous populations exhibiting isolation by distance \citep{white1968models}. When dispersal is limited and mating occurs locally, genetic differentiation naturally accumulates with geographic distance \citep{moritz1992evolutionary, irwin2005speciation, martins2013, de2017barriers}. If mating compatibility also depends on genetic similarity, reproductive barriers can emerge without ecological divergence, driven purely by spatial structure and the gradual accumulation of genetic differences beyond a compatibility threshold \citep{Manzo_Peliti_1994, kondrashov1999interactions, de2009global}. Heterogeneous habitats can further promote local adaptation and facilitate speciation \citep{lande1982rapid, rainey1998adaptive, schliewen2001genetic}, but classical theoretical studies often assume that divergent selection across environmental gradients drives diversification through competition or assortative mating acting directly on an ecological trait \citep{lewontin1968, kondrashov1998, Dieckmann_2003, Kopp2014, Lenormand2015}. Distinguishing whether ecological divergence initiated speciation or accumulated after reproductive isolation had already evolved therefore remains difficult.

To disentangle these alternative pathways, we develop a spatially explicit individual-based model that separates the genetic basis of ecological adaptation from that of reproductive compatibility. This separation allows us to examine whether ecological divergence precedes reproductive isolation or becomes pronounced only after reproductive isolation has begun to emerge under different ecological and mating conditions. Our approach builds on theoretical work showing that spatially restricted dispersal, local mating, and genetic compatibility can generate reproductive isolation in the absence of environmental heterogeneity \citep{Manzo_Peliti_1994, de2009global, martins2013, de2017barriers, costa2019,princepe2022diversity, freitas2022speciation}. 

Individuals in our model disperse locally and mate only with genetically compatible partners within a finite spatial neighborhood, generating isolation by distance and allowing reproductive isolation to arise through local mating and genetic compatibility alone. At the same time, a separate set of environmental loci determines the ecological phenotype, whose match to the local environment affects individual fitness but plays no direct role in reproductive compatibility. Individuals with phenotypes better adapted to the local environment have higher reproductive success, whereas mating compatibility is determined exclusively by genetic similarity at a distinct set of compatibility loci. Ecological adaptation therefore influences reproduction indirectly through fitness differences rather than by altering mate compatibility. In contrast to many models of ecological speciation along environmental gradients, in which ecological traits directly influence mate choice or competitive interactions \citep{Dieckmann_2003, Kopp2014, Lenormand2015}, ecological phenotype in our model is neither a mating cue nor a criterion for genetic compatibility. Consequently, the model excludes magic traits \citep{servedio2011magic}, assortative mating based on ecological phenotype, and reinforcement \citep{nosil2012ecological}.

Using this framework, we ask three questions. First, how does environmental selection modify diversification generated by spatial structure, local mating, and genetic compatibility? Second, does ecological divergence necessarily precede reproductive isolation, or can pronounced ecological differentiation emerge only after reproductive isolation has begun? And finally,  can bimodal phenotypic distributions across an environmental boundary arise through different evolutionary pathways?

By comparing homogeneous and heterogeneous environments across contrasting mating restrictions and selection strengths, we show that environmental selection can accelerate diversification and organize phenotypes around local optima without necessarily being the primary cause of reproductive isolation. Ecological differentiation and reproductive isolation do not follow a unique temporal sequence: ecological divergence may either (i) precede and contribute to its emergence, (ii) precede species formation without being necessary for it, or (iii) become pronounced only after reproductive subdivision has begun through spatial-genetic processes. Consequently, bimodal phenotypic patterns can arise through distinct evolutionary histories, cautioning against inferring the mechanism or causal sequence of speciation from phenotypic differentiation alone.

% %%%%%%%%%%%%%%%%%%%%%%%%%%%%%%%%%%%%%%%%%%%%%%%%%%%%
% %%%%%%%%%%%%%%%%%%%%%%%%%%%%%%%%%%%%%%%%%%%%%%%%%%%%
\section*{Methods}

We propose an individual-based model (IBM) on a $L_1 \times L_2$ lattice with reflective (non-periodic) boundaries to investigate speciation in sexually reproducing populations subject to natural selection and hard genetic compatibility constraints based on genetic similarity. Initially, $M_0$ individuals occupy the lattice randomly and, although typical population densities are small, multiple individuals can share the same site. The model explores how individuals adapt and diversify between distinct environments, even in the absence of physical barriers. Unlike models with fixed population sizes, here the population dynamics are self-regulated by local and global carrying capacities. Consequently, the total population size $M(t)$ is a dynamic variable regulated by ecological constraints, tending towards a steady state centered around $M_0$. 

Each individual $i$ carries two chromosomes: one for reproductive compatibility, denoted by $\gamma_i = (\gamma_i^1, \gamma_i^2, ..., \gamma_i^B)$, and one for environmental compatibility, $\rho_i = (\rho_i^1, \rho_i^2, ..., \rho_i^C)$, with lengths $B$ and $C$, respectively. The chromosomes are represented by binary sequences, so that $\gamma_i^k$ and $\rho_i^k$ can only assume the values $0$ or $1$. The ecological phenotype $P_i$ is determined by the additive effects of the environmental loci:
\begin{equation}
	P_i = \frac{1}{C} \sum_{k=1}^C  \rho_i^k
	\label{eq::phenotype}
\end{equation}

The environment exerts selective pressure based on the match between an individual's phenotype and the local environmental optimum, $P_E$. The individual fitness, $w_i$, is computed using a Gaussian function:
\begin{equation}
w_i = e^\frac{-(P_i - P_E)^2}{2\sigma_e^2},
\label{eq::w}
\end{equation}
where $\sigma_e$ represents the selection width; smaller values of $\sigma_e$ imply a narrower fitness peak and thus stronger selective pressure.

\begin{figure}[htb!]
\centering
\includegraphics[alt={Representation of the model showing the population and the recombination scheme during reproduction.},width=\textwidth]{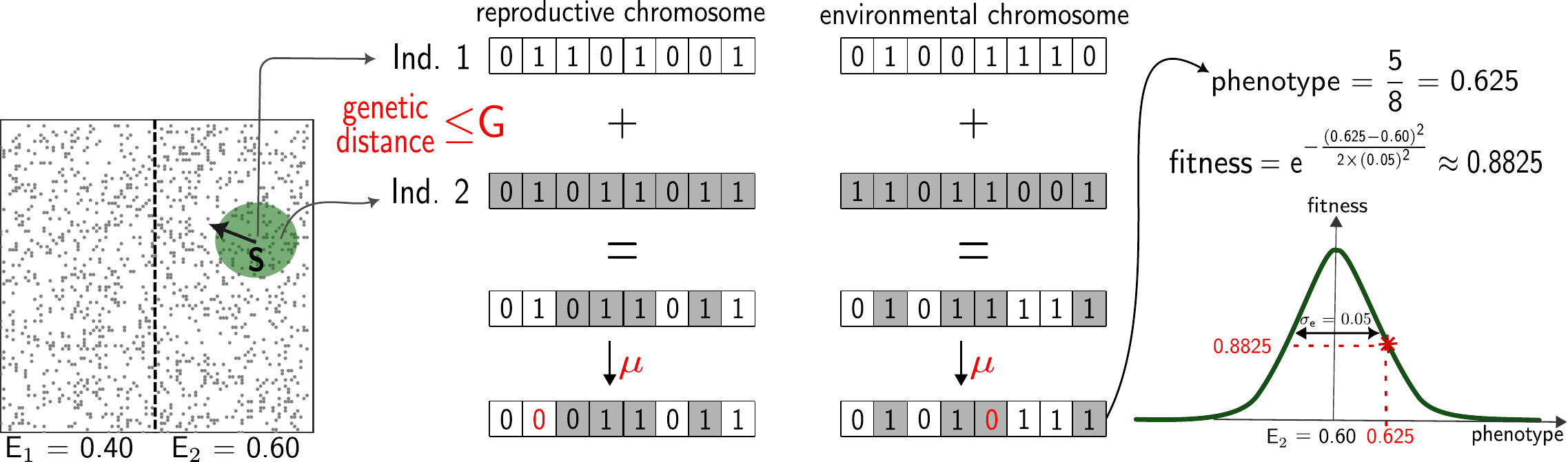}
    \caption{Representation of the model: the lattice with $M$ individuals is divided into two environments. An individual is selected and chooses a genetically compatible mating partner within a radius $S$, such that $D_{ij} \leq G$. The offspring inherits a random combination of reproductive loci from both parents. Each locus may mutate with a given probability, flipping the value of the binary allele. The phenotype is determined by the sum of the loci in the environmental-chromosome, which determines the offspring's fitness. The new generation of individuals replaces the old one, and the process is repeated over time, allowing the population to evolve under the influence of mutation, selection, and reproductive constraints.}
    \label{fig:model}
\end{figure}

We consider two environmental scenarios. The first is a homogeneous environment with a single ecological optimum, which works as a null model. The second is a heterogeneous environment with two distinct optima. In this case, the lattice is divided along the horizontal axis $L_1$ into two equal regions: $E_1$ (covering $1 \leq x \leq L_1/2$) and $E_2$ (covering $L_1/2 < x \leq L_1$), with respective optima $P_{E_1}$ and $P_{E_2}$ -- see Fig.~\ref{fig:model}. 

Focal individuals can select mating partners across both environments, including across their boundaries, ensuring that the entire space is connected without any physical barrier. However, mating partners are chosen only within a mating radius $S$ centered on the focal individual. This neighborhood is defined by the Euclidean distance between the centroids of the lattice sites. Specifically, all individuals located within cells whose centers fall within the distance $S$ from the center of the focal individual’s cell are considered potential mating partners. Moreover, mating is only possible if the genetic distance between the reproductive chromosomes of the mating pair is sufficiently small. The genetic distance $D_{ij}$ between two individuals, $i$ and $j$, is defined by
\begin{equation}
 D_{ij} = \sum_k |\gamma_i^k - \gamma_j^k|,
 \label{eq::Dij}
\end{equation}
and corresponds to the number of differing loci between their reproductive chromosomes (see Fig.~\ref{fig:model}).

Mating occurs only if $D_{ij}$ does not exceed the compatibility threshold $G$, defined as the maximum number of differing loci allowed for successful reproduction. This introduces a genetic compatibility filter acting within a spatial neighborhood, effectively implementing a distance-dependent mechanism of hard genetic compatibility constraint. Thus, an individual $i$ can reproduce only if it finds genetically compatible mating partners within its mating neighborhood $S$, differing in no more than $G$ loci across their reproductive chromosomes. Among these potential mating partners, one is chosen at random, without any feedback from the ecological phenotype. However, if no genetically compatible partners are found within the radius $S$, the candidate neighbor list remains empty, and the focal individual's reproductive attempt fails, producing no offspring for that generation.

We implement sexual reproduction with free recombination, where each locus of the offspring's chromosomes is inherited independently from either parent with equal probability. Following recombination, each locus may undergo a mutation (from 0 to 1 or from 1 to 0)  with probability $\mu$ per locus. This mechanism ensures sufficient genetic variation for adaptation within the simulated timescales. 

The resulting offspring is placed at a randomly selected site within the reproduction radius $S$ of the focal parent. The overall model workflow is illustrated in Fig.~\ref{fig:model}, which summarizes the key components of the process: the radius $S$, the chromosomal structure, and the fitness calculation  that  governs reproductive dynamics between individuals.

Versions of this model have previously been considered in a neutral scenario \citep{de2009global}, where all individuals had an equal probability of reproducing regardless of genotype. Evolution under those conditions, governed only by spatial proximity and genetic compatibility, can already lead to speciation. In contrast, our extended model incorporates environmental selection, making reproduction dependent on individual fitness. To bridge these two versions, we define the probability $m_i$ that an individual $i$ fails to reproduce (or, similarly, does not survive for reproduction) due to poor fitness as:
\begin{equation}
m_i = 1 - (1 - m')w_i,
\label{eq:q}
\end{equation}
where $m'$ is the baseline probability of reproductive failure and $w_i$ is the individual's fitness -- see Eq.~(\ref{eq::w}). The probability of successful reproduction is thus $1 - m_i$. In a neutral scenario ($\sigma_e \to \infty$), $w_i \to 1$ and $m_i = m'$ for all individuals. Conversely, as $w_i \to 0$, $m_i \to 1$, effectively preventing the individual from reproducing. This formulation increases the reproductive success of well-adapted individuals (up to the value $1 - m'$) while penalizing poorly adapted ones. Following previous work \citep{de2009global}, we fix $m' = 0.37$, which, mathematically,  corresponds to the probability that an individual is not selected for reproduction in $M$ trials with replacement from a population of size $M$, \textit{i.e.} $(1 - 1/M)^M \approx e^{-1}$. This choice controls baseline demographic stochasticity and does not qualitatively affect model behavior across the explored parameter ranges.

The model operates with non-overlapping generations, where the population for each subsequent time step is determined by a stochastic process incorporating compensatory recruitment. When the population $M(t)$ falls below the carrying capacity $M_0$, the per-capita fecundity potential increases according to $\eta = \lceil M_0/M(t) \rceil$. This mechanism captures an ecological scenario where reduced competition for resources in sparse populations allows for  higher reproductive success. The parameter $\eta$ sets the maximum offspring allowance for a reproductive event. Near demographic equilibrium ($M(t) \approx M_0$), each focal individual that found a mating pair is allowed to leave a single offspring ($\eta = 1$), whereas a drop in global density increases this allowance ($\eta > 1$).

 Therefore, successful reproduction requires the focal individual to pass the probability of surviving for reproduction (Eq. \ref{eq:q}) and then finding a genetically compatible partner within its mating radius $S$. If these conditions are met, primary offspring production is guaranteed. For the survival of extra offspring ($\eta > 1$)  we included a condition on local competition: the local spatial density -- defined as the number of neighbors $k_i$ within the mating radius $S$ of focal individual $i$ -- must be below a saturation threshold $K_{local}$. We derive this threshold as the expected number of individuals in a saturated neighborhood:
\begin{equation}
K_{local} = \frac{M_0}{L_1 L_2} \pi S^2,
\end{equation}
If $k_i \ge K_{local}$, the local neighborhood is saturated, and the mechanism blocks survival of extra offspring. Additional offspring can only be produced if space is available ($k_i < K_{local}$) and their survival remains probabilistic and governed by the focal individual's fitness, with a success probability of $1 - m_i$ (Eq. \ref{eq:q}). Consequently, a well-adapted individual in a sparse region may produce up to $\eta$ offspring, whereas a poorly adapted individual or one located in a crowded hub may produce only one or none. This mechanism ensures that population growth respects ecological carrying capacity, spatial constraints, and natural selection. In the Supplementary Material (SM), we present a sensitivity analysis of species richness as a function of the mating range $S$ (Fig S1) for both homogeneous and heterogeneous scenarios).

To characterize the population structure, species are identified as clusters of individuals linked by potential gene flow. Formally, we represent the population as a network where individuals are nodes and edges denote reproductive compatibility. In this framework, an adjacency matrix is defined such that $A_{ij}=1$ if $D_{ij} \leq G$, and $A_{ij}=0$ otherwise. Species then correspond to the connected components of this network, which we treat as operational units defined by potential gene flow rather than by phenotypic or ecological criteria. Mechanistically, while $G$ acts as a strict bound for immediate pairwise mating, macro-level species emerge dynamically as interconnected networks of individuals. The total genetic distance between two members of the same species can exceed $G$, since intermediate interbreeding individuals can connect them. Under this definition, the threshold $G$ functions as a pre-zygotic barrier based on genetic compatibility, while environmental selection -- Eq.~(\ref{eq::w}) -- acts as a filter for locally maladapted individuals, reducing the reproductive contribution of offspring with an intermediate (out-of-optima) phenotype and making them unlikely to survive to reproduce. These phenotypically intermediate individuals can be generated via matings between individuals (parents) from the same species that have different phenotypes. In the SM, we present a sensitivity analysis of species richness as a function of the threshold $G$ under different selection strengths (Fig S2).

\section*{Numerical Simulations}

To investigate how reproductive isolation can operate independently of ecological divergence in spatially continuous environments, we conducted a series of numerical experiments on a lattice of size $L_1 = L_2 = 100$. The system was initially populated by $M_0 = 1,300$ individuals, resulting in a global density of $0.13$ individuals per square unit when the population is at its carrying capacity. Spatial interactions were governed by a mating radius $S = 6$, which provides an average of $14.7$ potential mating partners per individual ($\pi S^2 \times 0.13$) when the population is at its carrying capacity.

The genetic architecture in our model consists of two independent modules: a reproductive chromosome with $B = 1,500$ loci and an environmental-chromosome with $C = 100$ loci. Both are subject to a mutation rate $\mu = 2.5 \times 10^{-4}$ per locus per generation. This results in a total mutational input of approximately 520 mutations per generation for the initial population
($M_0 \times (B + C) \times \mu = 1,300 \times 1,600 \times 2.5 \times 10^{-4}$). This mutational input provides sufficient genetic variation for adaptation and speciation to be observed within a computationally feasible timescale ($t = 5{,}000$).

Within this framework, we explored the permeability of genetic boundaries by varying the mating compatibility threshold $G$, contrasting a restrictive regime ($G = 0.05B = 75$ loci) with a permissive regime ($G = 0.30B = 450$ loci). These values of $G$ were chosen because in a neutral selection scenario ($\sigma_e \approx \infty$), speciation is possible for the restrictive value ($G = 0.05B$), but not for the permissive one ($G = 0.30B$) -- see Fig. S4 in the SM. Concurrently, the strength of natural selection was modulated through the selection width $\sigma_e$, comparing strong selection ($\sigma_e = 0.05$) with weak selection ($\sigma_e = 0.40$).

By cross-combining these extreme values, we define four evolutionary regimes that represent the boundary conditions of the model: restrictive mating and strong selection (RS; $G = 0.05B$, $\sigma_e = 0.05$), restrictive mating and weak selection (RW; $G = 0.05B$, $\sigma_e = 0.40$), permissive mating and strong selection (PS; $G = 0.30B$, $\sigma_e = 0.05$), and permissive mating and weak selection (PW; $G = 0.30B$, $\sigma_e = 0.40$). These four extremes allow a clear characterization of the interplay between spatial structure and selective pressure throughout our analysis.

To illustrate the impact of these selection regimes, consider an individual with a phenotypic deviation of 0.10 from the local optimum; according to Eq.~(\ref{eq::w}), its fitness is reduced to
\[
w = \exp\left( -\frac{0.10^2}{2 \cdot 0.05^2} \right) \approx 0.135
\]
under strong selection, compared to $w \approx 0.969$ under weak selection ($\sigma_e = 0.40$). These contrasting regimes allow us to evaluate the conditions under which phenotypic divergence is maintained or eroded despite ongoing gene flow.

To ensure that any observed divergence arises from the evolutionary dynamics rather than initial bias, all individuals were initialized with non-identical environmental-chromosomes, drawn in a way to obtain a uniform distribution of phenotypes in the interval $\left[ 0.3, 0.7 \right]$. Regarding reproductive compatibility, individuals start with exactly the same reproductive chromosome (all loci with ``0''), therefore being of the same species and under maximum compatibility with each other. We allowed the populations to evolve for $5{,}000$ generations under two environmental scenarios: a homogeneous scenario with a single optimal phenotype of $P_E = 0.50$, and a heterogeneous scenario with two divergent optima $P_{E_1} = 0.40$ and $P_{E_2} = 0.60$. The $5{,}000$-generation simulation window was sufficient to characterize the main diversification dynamics across the investigated parameter combinations. Additional simulations extending to $10{,}000$ generations were performed for the RW regime to assess its long-term non-equilibrium phenotypic dynamics.

\section*{Results}

We first consider the homogeneous environment, which serves as a baseline for evaluating how spatial structure, local mating, and genetic compatibility can generate diversification in the absence of ecological differentiation. Next, we move to the heterogeneous environment, where we investigate how divergent selection modifies spatial-genetic diversification across ecological boundaries.

\subsection*{Homogeneous environment}
\begin{figure}[!htpb]
\centering
  \includegraphics[alt={Diversification dynamics in homogeneous environments showing populations and phenotipic distributions. },width=0.8\textwidth]{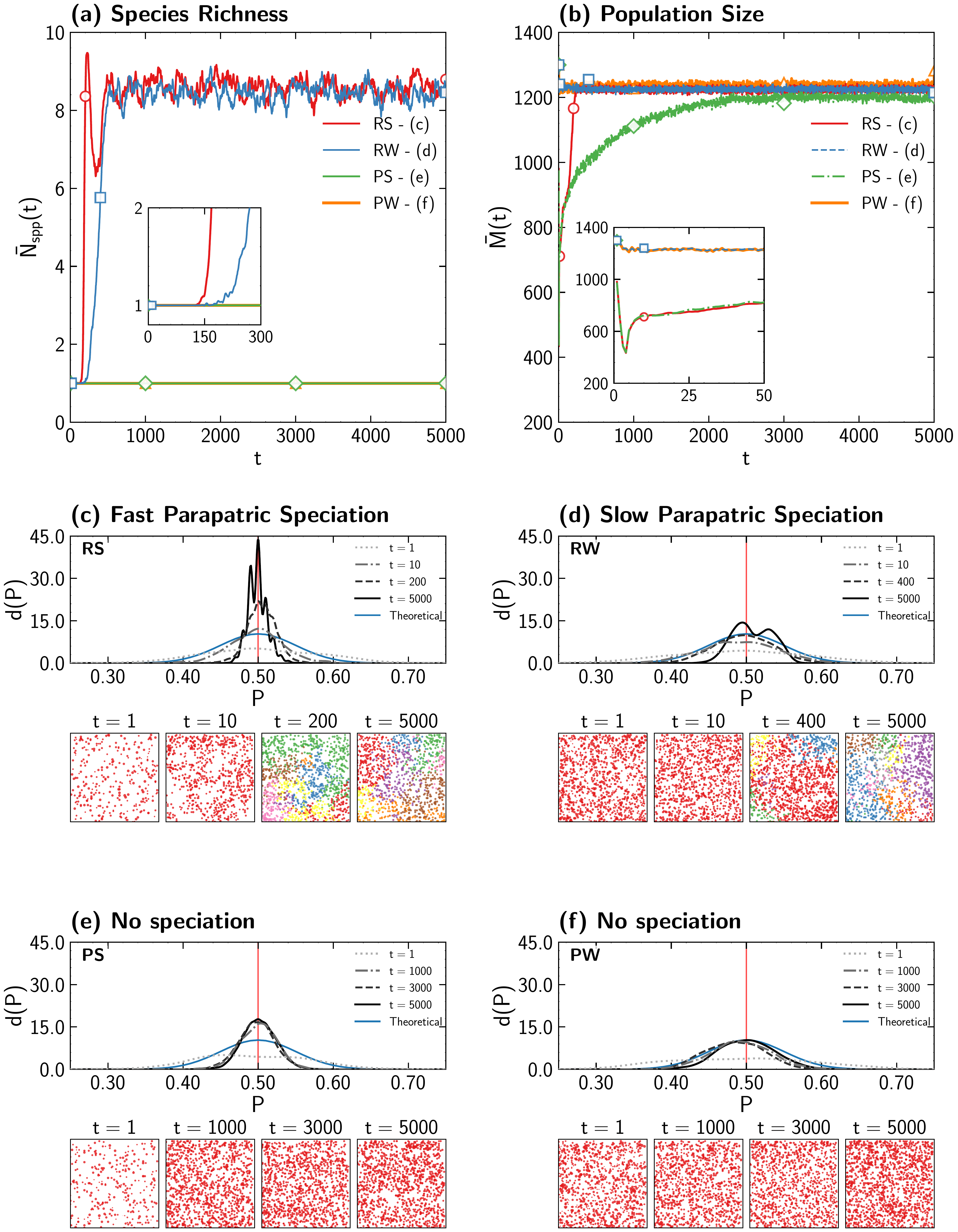}
    \caption{Diversification dynamics in homogeneous environments. (a) Evolution of species richness, $N_{spp}(t)$, and (b) total population size, $M(t)$, averaged over 50 simulations. (c--f) Phenotypic density distributions $d(P)$ and corresponding spatial organization at selected times (shown in (a,b) by symbols) for a representative replicate and different parameter combinations: (c) RS ($G = 0.05B$, $\sigma_e = 0.05$) -- fast parapatric speciation under restrictive mating and strong selection; (d) RW ($G = 0.05B$, $\sigma_e = 0.40$) -- slow parapatric speciation under restrictive mating and weak selection; (e) PS ($G = 0.30B$, $\sigma_e = 0.05$) -- no speciation under permissive mating and strong selection; (f) PW ($G = 0.30B$, $\sigma_e = 0.40$) -- no speciation under permissive mating and weak selection. Dots represent individuals, with their sizes proportional to the ecological phenotype $P$. Different colors represent species. In (c), (d), (e), and (f), the red vertical line indicates the local environmental optimum, $P_E=0.50$; the blue curve indicates the theoretical neutral phenotypic distribution expected for ecological loci with free recombination, in the absence of selection.}
    \label{fig:results_homogeneous}
\end{figure}

We first examine evolution on a homogeneous landscape with a single ecological optimum, $P_E=0.50$, across the entire lattice. In the absence of ecological heterogeneity, any diversification must arise from spatial structure and reproductive constraints rather than niche differentiation. Population size reflects the balance between reproductive success, natural selection, and local density-dependent competition. The temporal dynamics of species richness $N_{spp}(t)$ and total population size $M(t)$ are shown in Fig.~\ref{fig:results_homogeneous}(a,b), averaged over 50 independent realizations.

No extinction occurred in any replicate across the explored parameter space. While these averages capture the overall temporal trends, the resulting phenotypic distributions and spatial organization are illustrated using a single representative simulation. The symbols in Fig.~\ref{fig:results_homogeneous}(a,b) mark four representative stages of the dynamics: (i) the initial response to selection, (ii) an early transient, (iii) an intermediate stage, and (iv) the long-time state at $t=5{,}000$. The corresponding phenotypic distributions and spatial configurations are shown in Fig.~\ref{fig:results_homogeneous}(c--f).

As shown in Fig.~\ref{fig:results_homogeneous}(a), speciation occurs exclusively under the restrictive mating regime ($G=0.05B$ -- red and blue lines), irrespective of selection strength (RS and RW scenarios). Species richness $N_{spp}$ rapidly increases and stabilizes at approximately 8--10 species within the first $1{,}000$ generations. In contrast, under the permissive mating regime ($G=0.30B$, PS and PW), the population remains a single species ($N_{spp}=1$) throughout the simulation. Thus, without a sufficiently restrictive genetic compatibility threshold, spatial structure and local mating are insufficient to overcome the homogenizing effect of gene flow.

Selection strength nevertheless strongly affects the timing of diversification and the associated demographic dynamics. Under restrictive mating ($G=0.05B$), speciation across the continuous spatial landscape occurs more rapidly under strong selection (RS, $\sigma_e=0.05$ -- see Fig \ref{fig:results_homogeneous}(a) red line) than under weak selection (RW, $\sigma_e=0.40$ -- see Fig \ref{fig:results_homogeneous}(a) blue line) or neutral selection ($\sigma_e\to\infty$; Fig.~S4(c)). Under RS, the replicate-averaged global phenotypic variance declines sharply from the beginning of the simulation (Fig.~S5(a) in the  SM), and the phenotypic distribution of the representative simulation is already strongly concentrated around the environmental optimum by $t=100$ (Fig.~\ref{fig:results_homogeneous}(c)). Mean species richness begins to increase later, at approximately $t\approx150$ (inset in Fig.~\ref{fig:results_homogeneous}(a) -- red line).

Strong selection initially imposes a high demographic cost, sharply reducing the mean population size $\bar{M}(t)$ because phenotypes furthest from the optimum experience high reproductive failure rates ($m_i$) (Fig.~\ref{fig:results_homogeneous}(b), red and green lines). Under strong selection, however, demographic recovery is substantially faster under restrictive than under permissive mating (Fig.~\ref{fig:results_homogeneous}(b), RS and PS -- red and green lines, respectively). In RS, the rapid concentration of phenotypes around the fitness optimum, $P_E=0.50$, is accompanied by an increase in global mean fitness toward one (Fig.~S5(b), red line). As mean fitness increases, reproductive failure declines, allowing compensatory recruitment to restore the population toward its carrying capacity (Fig.~\ref{fig:results_homogeneous}(b) -- red line). Under strong selection but permissive mating (PS), reproductive isolation does not evolve; as a consequence, global phenotypic variance remains higher than under RS (Fig.~S5, green line). The demographic recovery under PS takes approximately ten times longer than under RS (Fig.~\ref{fig:results_homogeneous}(b), green line compared to the red line).

Although selection strength has little effect on long-term species richness within each mating regime, the phenotypic density distributions confirm that strong selection produces (Fig.~\ref{fig:results_homogeneous}(c, e)) a narrower distribution than weak selection (Fig.~\ref{fig:results_homogeneous}(d, f)). Additionally, under strong selection the phenotypic variance became smaller (Fig.~S5(a) - red and green lines) than the expected by the neutral case ($Var(P)=0.0025$) earlier than the weak selection cases (blue and orange lines). For the restrictive mating cases, it is also possible do detect that phenotypic variation decreases more after speciation has started (Fig.~S5(a) -- red and blue lines). For instance, under RW, mean species richness begins to increase before $t\approx300$, whereas the replicate-averaged global phenotypic variance remains comparatively high during the early dynamics (close to the neutral expected variance) and declines only thereafter (Fig.~S5(a) -- blue line). Consistently, in the representative simulation, a clear phenotypic response to weak selection -- indicated by a distribution narrower than the theoretical neutral expectation, is evident only by $t=400$ (Fig.~\ref{fig:results_homogeneous}(d)).

As a methodological control, we compared these results with a fully permissive scenario ($G=B$), in which the genetic compatibility restriction is effectively removed and reproductive isolation does not emerge (see Fig.~S3(a,b) in the SM). As detected in the under restrictive and permissive mating scenarios (Fig.~\ref{fig:results_homogeneous}(c, e)), strong selection also narrows the neutral baseline distribution under strong selection (Fig.~S3(a)). However, secondary peaks that emerge on top of the expected population-wide phenotypic distribution under restrictive mating (Fig.~\ref{fig:results_homogeneous}(c,d)) are not observed when no species emerge (Fig.~\ref{fig:results_homogeneous}(e,f) and Fig.~S3(a,b). Individuals belonging to different reproductively isolated species cluster around slightly different phenotypic means, generating secondary peaks that are absent when the compatibility restriction is removed or prevents species to form.

Together, these results show that, in the homogeneous environment, the genetic compatibility threshold determines whether speciation occurs, whereas environmental selection primarily affects the timing of diversification, the population-wide phenotypic distribution, and demographic recovery. The observed species richness ($\bar{N}_{spp}\approx8$--$10$) is lower than the theoretical limit predicted for finite populations with infinitely long genomes under mean-field mating. A detailed comparison between our simulation results and the analytical Derrida-Higgs model is provided in Section~S6 of the SM.

\subsection*{Heterogeneous environment}

We next investigate how ecological heterogeneity modifies the diversification dynamics established in the homogeneous environment. To this end, we simulated a landscape composed of two adjacent environments with distinct phenotypic optima, $P_{E_1}=0.40$ and $P_{E_2}=0.60$. Because no physical barrier separates the two habitats, individuals can disperse and reproduce across the environmental boundary, generating a parapatric setting in which divergent environmental selection and spatial-genetic processes act simultaneously.

\begin{figure}[!htpb]
\centering
  \includegraphics[alt={Diversification dynamics in heterogeneous environments showing populations and phenotipic distributions.},width=0.8\textwidth]{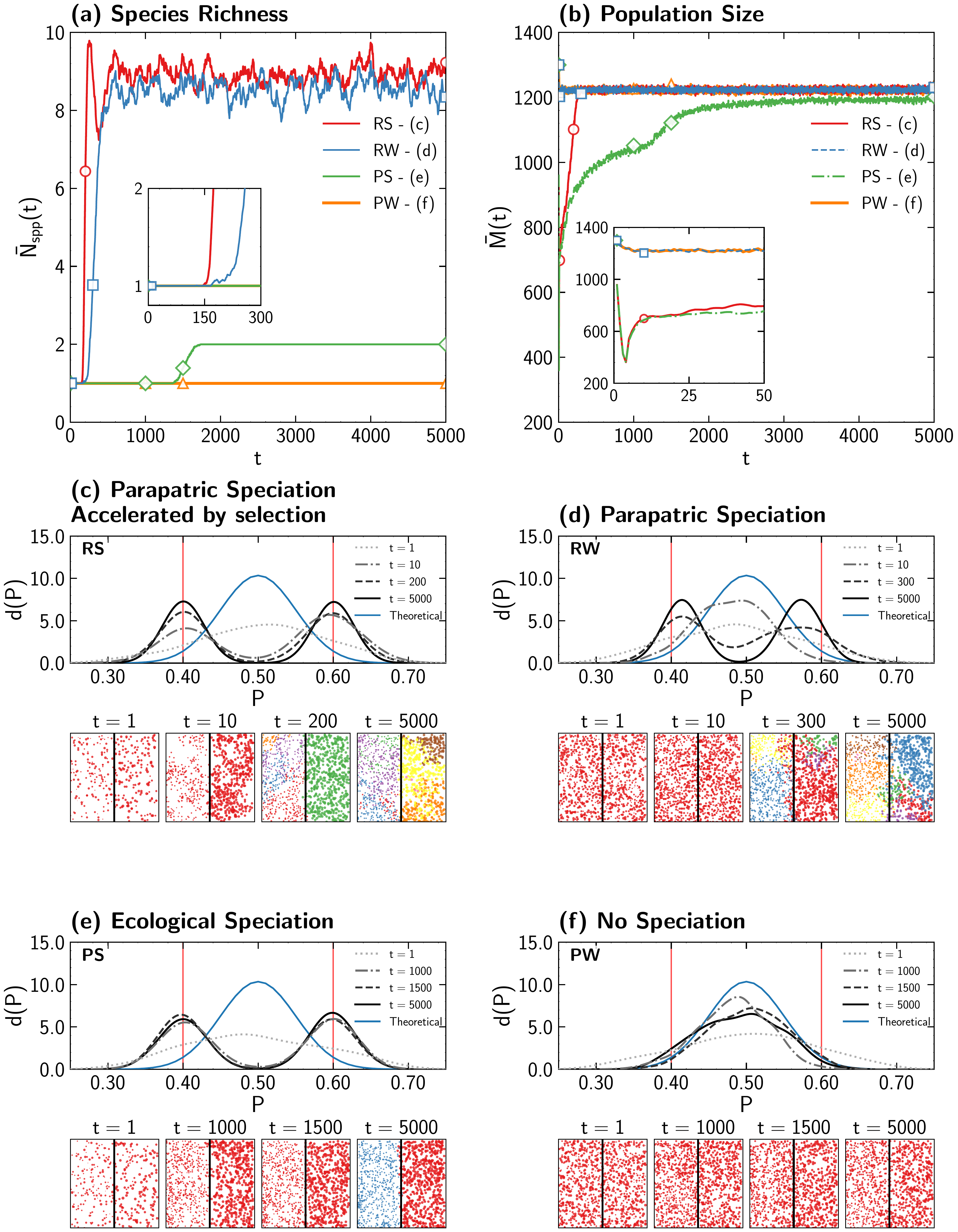}
   \caption{Diversification dynamics in heterogeneous environments. (a) Evolution of species richness, $N_{spp}(t)$, and (b) total population size, $M(t)$, averaged over 50 simulations. (c--f) Phenotypic density distributions $d(P)$ and corresponding spatial organization at selected times (shown in (a,b) by symbols) for a representative replicate and different parameter combinations: (c) RS ($G=0.05B$, $\sigma_e=0.05$) -- parapatric speciation accelerated by strong selection under restrictive mating and strong selection; (d) RW ($G=0.05B$, $\sigma_e=0.40$) -- parapatric speciation under restrictive mating and weak selection; (e) PS ($G=0.30B$, $\sigma_e=0.05$) -- ecological speciation under permissive mating and strong selection; and (f) PW ($G=0.30B$, $\sigma_e=0.40$) -- no speciation under permissive mating and weak selection. Dots represent individuals, with their sizes proportional to the ecological phenotype $P$. Different colors represent species. In (c), (d), (e), and (f), the red vertical lines indicate the local environmental optima, $P_{E_1}=0.40$ and $P_{E_2}=0.60$; the blue curve indicates the theoretical neutral phenotypic distribution expected for freely recombining ecological loci in the absence of selection.}
    \label{fig:results_heterogeneous}
\end{figure}

The heterogeneous environment allows us to compare diversification under restrictive and permissive mating across contrasting selection regimes. Figure~\ref{fig:results_heterogeneous}(a,b) shows the temporal evolution of species richness $N_{spp}(t)$ and total population size $M(t)$, averaged over 50 independent simulations. As in the homogeneous environment, no extinction occurred in any replicate across the explored parameter space and  we highlight four representative stages of the dynamics: (i) the initial response to selection, (ii) an early transient, (iii) an intermediate stage, and (iv) the long-time state at $t=5{,}000$. Corresponding phenotypic distributions and spatial configurations from a representative simulation are shown in Fig.~\ref{fig:results_heterogeneous}(c--f).

We first examine the restrictive mating regime ($G=0.05B$), under which diversification occurs with both strong and weak selection (RS and RW, Fig.~\ref{fig:results_heterogeneous}(a) -- red and blue lines). The temporal relationship between ecological differentiation and reproductive isolation, however, differs markedly between these selective regimes. Under strong selection (RS, $\sigma_e=0.05$), the replicate-averaged global phenotypic variance rises sharply from the beginning of the simulation as the population separates toward the two environmental optima (Fig.~S6(a) in the SM -- red line and a representative simulation in Fig.~\ref{fig:results_heterogeneous}(c)). This phenotypic response precedes the increase in mean species richness, which begins at approximately $t=150$ (inset in Fig.~\ref{fig:results_heterogeneous}(a)). The representative simulation illustrates how rapidly this response occurs: by $t=10$, a bimodal phenotypic distribution aligned with the two local optima is already evident, although the population still belongs to a single reproductive species (Fig.~\ref{fig:results_heterogeneous}(c)). Multiple reproductively isolated species emerge only later through the combined effects of spatial structure, local mating, and restrictive genetic compatibility.

During this transition, strong selection reduces the persistence of maladapted intermediate phenotypes near the environmental boundary, while the restrictive compatibility threshold prevents reproduction between individuals with sufficiently divergent reproductive genotypes. Consistently, successful mating pairs near the boundary are more strongly concentrated at small phenotypic differences under strong selection (Fig. S7(a)). Thus, under RS, ecological differentiation precedes reproductive isolation. 

Strong selection also imposes an initial demographic cost because individuals with phenotypes far from their local environmental optimum experience high reproductive failure rates. Under RS, however, the rapid separation of the population toward the local environmental optima ($P_{E_1}=0.40$ and $P_{E_2}=0.60$) increases global mean fitness and reduces reproductive failure, allowing the total population size to recover toward its carrying capacity (Fig.~S6(b); Fig.~\ref{fig:results_heterogeneous}(b) -- red line in both figures). At this point, the global phenotypic variance also increases (Fig.~S6(b) -- red line), but it does not indicate increasing maladaptation; rather, it primarily reflects the growing separation between the two environment-associated phenotypic modes.

In contrast, under restrictive mating and weak selection (RW, $\sigma_e=0.40$), the temporal sequence is reversed. Mean species richness begins to increase before $t\approx300$, whereas the replicate-averaged global phenotypic variance changes only gradually during the early dynamics (Fig.~\ref{fig:results_heterogeneous}(a); Fig.~S6(a) -- blue line in both figures). Consistently, in the representative simulation, the population already contains multiple reproductively isolated species by $t=300$, while the phenotypic distribution remains broad and only partially structured rather than forming two clearly resolved environment-associated peaks (Fig.~\ref{fig:results_heterogeneous}(d)). Thus, reproductive isolation begins to emerge before pronounced ecological differentiation becomes evident. Once reproductively isolated species have formed, weak local selection gradually shifts their species-specific phenotypic means toward the corresponding environmental optima. The long-term bimodal pattern therefore develops after genetic fragmentation rather than initiating it.

\begin{figure}[!htpb]
\centering
    \includegraphics[alt={Phenotypic distributions over time },width=0.9\textwidth]{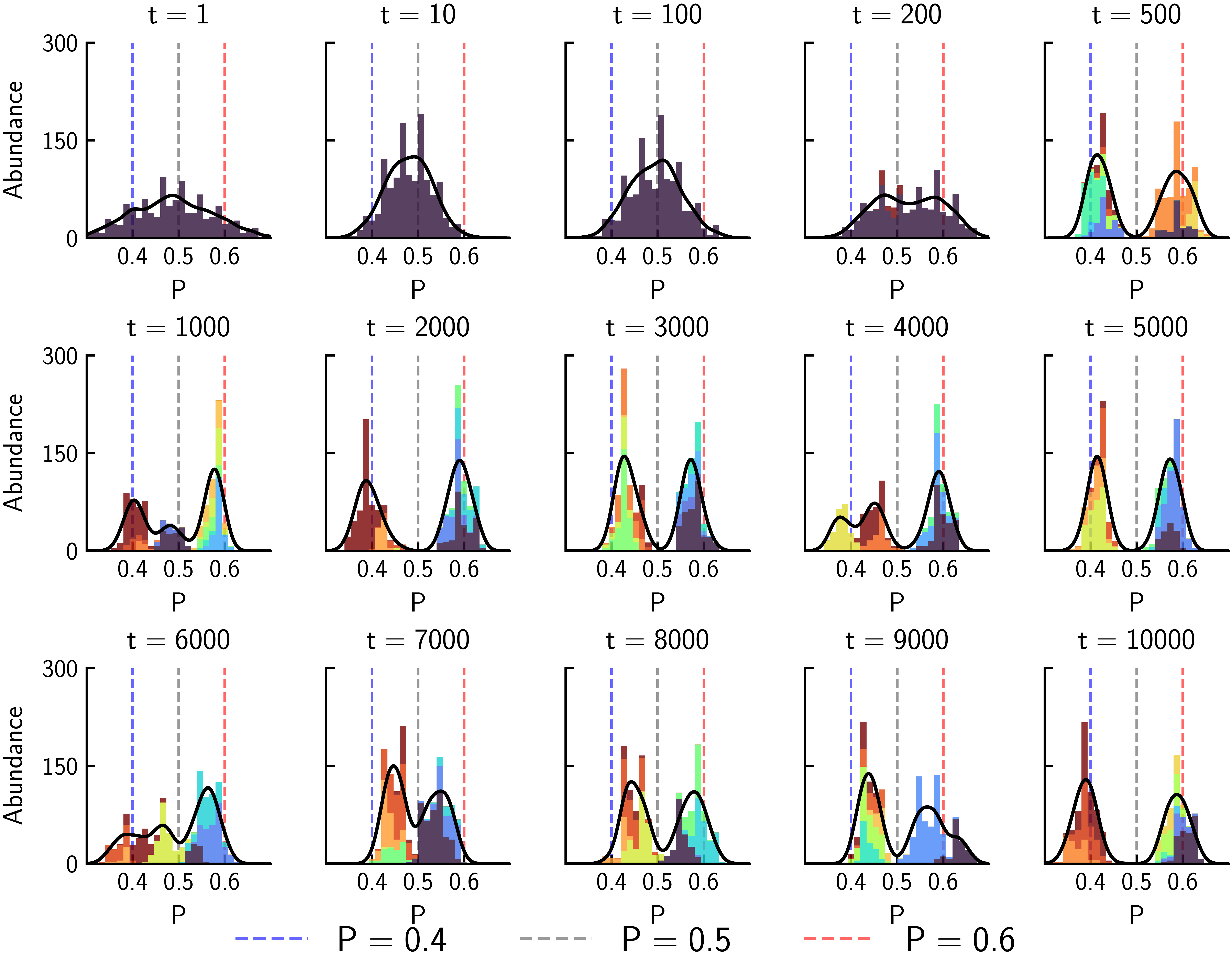}
    \caption{Phenotypic distributions over time for a representative simulation under weak selection ($\sigma_e=0.40$) and restrictive mating ($G=0.05B$). Species-specific phenotypic peaks shift persistently between the local environmental optima ($P=0.40$ and $P=0.60$, blue and red dashed lines respectively) and the neutral phenotypic expectation ($P=0.50$, gray dashed line). Colored distributions represent different reproductively isolated species. Colors are assigned independently at each time point and therefore do not track species lineages through time.}
    \label{fig:G5-2complet}
\end{figure}

Analysis of the long-term phenotypic dynamics further shows that, under restrictive mating and weak selection, the phenotypic distributions do not converge to stable positions at the environmental optima. As shown in Fig.~\ref{fig:G5-2complet}, species-specific phenotypic peaks continue to shift over time, even up to $t=10{,}000$ generations, while reproductive isolation remains intact. The system therefore persists in a non-equilibrium regime rather than approaching a stationary bimodal distribution. Consequently, a single temporal snapshot may capture phenotypic peaks displaced from the local optima, although these positions represent transient stages of the long-term dynamics rather than a stable configuration. Despite these fluctuations, the population does not undergo demographic decline. Because selection is weak, individuals retain high average fitness even when species-specific phenotypic means deviate from the environmental optima, and the total population size, $M(t)$, remains close to carrying capacity throughout the simulation (Fig.~\ref{fig:results_heterogeneous}(b) -- blue line). It is also possible to detect that the phenotypic variance has larger amplitude variations over time under the RW scenario (Fig.~S6(a) -- blue line) after species were formed).

To distinguish these dynamics from the effects of mutation and genetic drift alone, we compared them with a null model lacking genetic compatibility restrictions ($G=B$; Fig.~S3(c,d) in the SM). Under strong selection, the population forms two phenotypic modes near the local environmental optima ($P=0.40$ and $P=0.60$), whereas under weak selection it remains concentrated around a single intermediate mode (at $P=0.50$.) The bimodal distribution observed under RW therefore (see a representative simulation in Fig.~\ref{fig:results_heterogeneous}(d)) cannot be explained by mutation and genetic drift alone; it depends on the reproductive boundaries generated by restrictive genetic compatibility.

Diversification dynamics change qualitatively when mating becomes more permissive ($G=0.30B$) (Fig.~\ref{fig:results_heterogeneous}(a) -- green and orange lines). Under this regime, reproductive isolation evolves only when selection is strong. In the PS scenario ($\sigma_e=0.05$), mean species richness remains equal to one for approximately $1{,}400$ generations, while global phenotypic variance indicates the population has differentiated between environments since early dynamics (Fig.~\ref{fig:results_heterogeneous}(a); Fig.~S6(a) -- green line in both figures). In the representative simulation, two environment-associated phenotypic modes are already clearly visible by $t=1{,}000$, although all individuals still belong to a single reproductive species (Fig.~\ref{fig:results_heterogeneous}(e)). Mean species richness has a sharper increase around $t=1,400$--$1,500$. Ecological differentiation therefore precedes reproductive isolation in this regime.

Because the permissive compatibility threshold continues to allow reproduction across the environmental boundary, intermediate phenotypes are repeatedly produced. Their lower reproductive success under strong selection maintains global mean fitness below than the observed under RS and slows demographic recovery (Fig.~S6(b); Fig.~\ref{fig:results_heterogeneous}(b) -- green line in both figures). Strong selection nevertheless acts as an ecological filter against intermediate phenotypes, progressively reducing effective gene flow across the environmental boundary until two reproductively isolated groups emerge. Because ecological phenotype is neither a mating cue nor a criterion for genetic compatibility, this reproductive isolation does not result from phenotype-based assortative mating or reinforcement.

When permissive mating is combined with weak selection (PW, $\sigma_e=0.40$), intermediate phenotypes experience only small fitness penalties. Gene flow therefore remains sufficient to homogenize the population, and reproductive isolation does not emerge. Species richness remains equal to one (Fig.~S6(a) -- orange line), global phenotypic variance remains lower than under strong selection (Fig.~S6 -- orange line) and mean fitness remains high throughout the simulation (Fig.~S6 -- orange line).

\begin{figure}[!htpb]
\centering
\includegraphics[alt={Reproductive-genetic differentiation plotted as a function of geographic distance},width=0.9\textwidth]{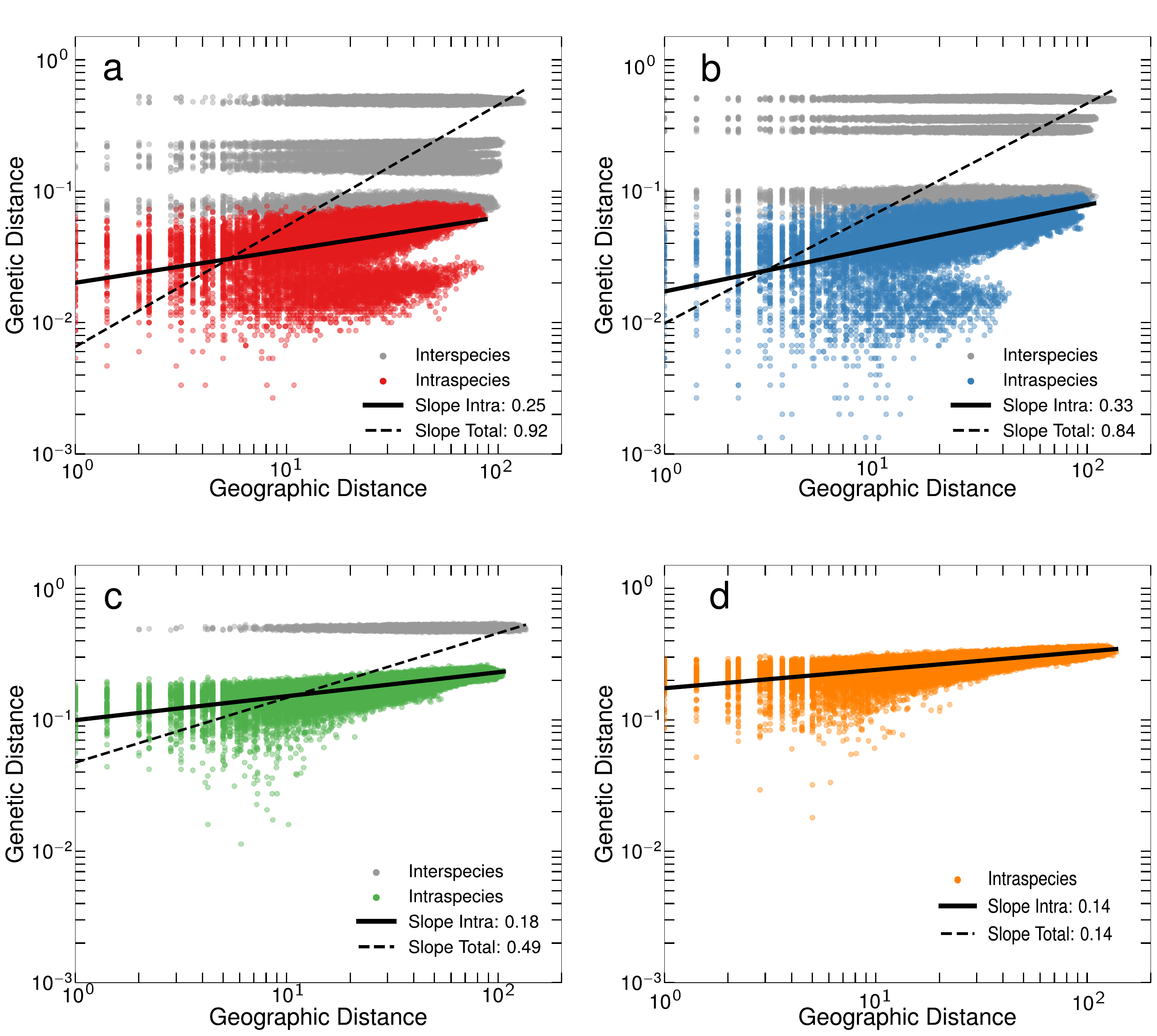}
\caption{Individual-level reproductive-genetic differentiation as a function of geographic distance at $t=5{,}000$, shown on log--log scales. Pairwise Euclidean geographic distance is plotted against the normalized Hamming distance of the reproductive chromosome, $D_{ij}/B$, where $D_{ij}$ is defined in Eq.~\ref{eq::Dij}. Colored dots represent pairs of individuals belonging to the same species, whereas gray dots represent pairs belonging to different species. (a,b) Restrictive mating ($G=0.05B$) under strong (a) and weak (b) selection, showing continuous isolation by distance within species and discrete reproductive-genetic differences between species. (c,d) Permissive mating ($G=0.30B$) under strong (c) and weak (d) selection; reproductive isolation occurs only under strong selection within the parameter combinations examined. Solid black lines show descriptive least-squares regressions fitted in log--log space to intraspecific comparisons, whereas dashed black lines show regressions fitted to all pairwise comparisons. The pooled regressions combine continuous within-species IBD with discrete between-species genetic differences. In (d), the solid and dashed regressions overlap because the population remains a single species. Data include all pairwise comparisons from one representative simulation for each parameter combination.}
\label{fig:hamming_ibd}
\end{figure}

To characterize spatial genetic structure, we analyzed individual-level isolation by distance (IBD) by relating pairwise Euclidean geographic distance to the normalized Hamming distance of the reproductive chromosome, $D_{ij}/B$, where $D_{ij}$ is defined in Eq.~\ref{eq::Dij}. Both quantities are shown on logarithmic scales in Fig.~\ref{fig:hamming_ibd}. Because mating and offspring dispersal are spatially restricted to the finite radius $S$, reproductive-genetic differences can accumulate with geographic separation even in the absence of physical barriers. The scatter plots include all pairwise comparisons from one representative simulation for each parameter combination at $t=5{,}000$.

Under restrictive mating ($G=0.05B$), the population separates into multiple connected components of the reproductive-compatibility network (Fig.~\ref{fig:hamming_ibd}(a,b)). Within species, normalized reproductive-genetic distance increases gradually with geographic separation, with log--log slopes of approximately $0.25$ under strong selection and $0.33$ under weak selection (solid line in Fig.~\ref{fig:hamming_ibd}(a,b)). Because species are defined by network connectivity rather than by direct compatibility between every pair of individuals, geographically distant members of the same species may have pairwise genetic distances exceeding the direct compatibility threshold ($G=0.05B$). Such individuals remain part of the same species provided that they are connected through an uninterrupted chain of pairwise compatible individuals.

Interspecific comparisons form approximately horizontal bands at characteristic levels of reproductive-genetic divergence. These bands reflect discrete divergence levels among disconnected components and show that the corresponding genetic discontinuities persist across a broad range of geographic distances. Because species assignment and the vertical axis are both derived from the reproductive chromosome, the separation between intra- and interspecific comparisons should be interpreted as a description of the spatial organization of the species identified by reproductive-genetic connectivity rather than as an independent test of reproductive isolation. When all pairwise comparisons are pooled, the descriptive regressions yield slopes of $0.92$ and $0.84$ under strong and weak selection, respectively (dashed line in Fig.~\ref{fig:hamming_ibd}(a,b)). These pooled slopes are steeper than the corresponding intraspecific slopes because they combine continuous within-species IBD with discrete genetic differences between species. They should therefore not be interpreted as conventional within-species IBD coefficients.

Under permissive mating ($G=0.30B$), reproductive isolation emerges only under strong selection within the parameter combinations examined. In the PS scenario ($\sigma_e=0.05$; Fig.~\ref{fig:hamming_ibd}(c)), the population eventually separates into two reproductive species. Within species, normalized reproductive-genetic distance increases with geographic separation with a slope of $0.18$, whereas the regression pooling all comparisons has a slope of $0.49$ (solid and dashed lines respectively in Fig.~\ref{fig:hamming_ibd}(c)). The larger pooled slope results from the inclusion of interspecific comparisons and therefore reflects both continuous within-species IBD and the genetic discontinuity between the two reproductive components. Under weak selection ($\sigma_e=0.40$; Fig.~\ref{fig:hamming_ibd}(d)), the intraspecific and pooled slopes are both $0.14$ because the population remains a single species. Thus, spatial genetic differentiation develops under permissive mating, but under weak selection it remains continuous and does not generate disconnected reproductive components within the simulated time interval.

To further examine how selection shapes successful reproduction near the ecological boundary, we analyzed the distribution of phenotypic differences, $\Delta P$, among pairs of individuals located within the boundary region, defined by the $x$-coordinate range $50\pm S$, that mated and produced an offspring. As shown in Fig.~S7 in the SM, selection strength alters the phenotypic composition of successful mating pairs near the interface. Under strong selection, intermediate phenotypes experience large fitness penalties in both environments and are therefore less likely to persist and reproduce. Successful mating pairs consequently become more strongly concentrated at small values of $\Delta P$. Under weak selection, intermediate phenotypes experience smaller fitness penalties and remain more abundant near the boundary, increasing the frequency of successful matings between phenotypically dissimilar individuals and producing flatter distributions of $\Delta P$. These patterns indicate greater phenotypic mixing among successful mating pairs near the ecological interface under weak selection, even when reproductive boundaries persist under restrictive mating.

Finally, to relate the emergent diversification patterns to their underlying genetic architecture, we examined differentiation across the environmental-chromosome in the heterogeneous landscape. Specifically, we compared mean environmental-chromosome distances between species occupying the same environment and species occupying different environments (Fig.~S8 in the SM) and examined loci that reached fixation or retained high levels of polymorphism (Fig.~S9 in the SM) in the long term.

Across both selection regimes, environmental-chromosome divergence is lower between species occupying the same environment and higher between species occupying different environments (see SM, Fig.~S8). Strong selection accentuates this contrast: divergence between species occupying the same environment is lower than under weak selection, whereas divergence between species occupying different environments is greater. Species occupying the same environment also share fixed alleles at a larger number of loci under strong selection (see SM, Fig.~S9 and Fig.~S10 in the SM) when analyzed in the long term. These patterns are consistent with stronger environment-specific sorting and parallel fixation of ecological alleles around the two local environmental optima.

%%%%%%%%%%%%%%%%%%%%%%%%%%%%%%%%%%%%%%%%%%%%%%%%%%%%%%%%%%%%%%%%%%%%%%%%%%%%%%%%%%%%%
\section*{Discussion}

Phenotypic divergence across environmental boundaries is frequently interpreted as evidence of divergent environmental selection driving speciation, particularly when bimodal trait distributions coincide with habitat differences and reduced gene flow \citep{rundle2005ecological, schluter2009evidence, nosil2012ecological, smadja2011framework}. Our results show, however, that similar phenotypic outcomes can emerge through distinct evolutionary histories. Under permissive mating and strong selection (PS), ecological differentiation indeed reduced effective reproductive exchange across the environmental boundary and preceded the emergence of reproductive isolation. Under restrictive mating and weak selection (RW), however, reproductive subdivision began while the phenotypic distribution remained broad, and pronounced environment-associated differentiation developed only afterward. Under restrictive mating and strong selection (RS), ecological differentiation also preceded species formation, but divergent selection was not necessary for reproductive isolation since the same restrictive mating and spatial-genetic processes produced species under weak selection in a homogeneous environment. Ecological differentiation may therefore contribute to reproductive isolation, emerge after reproductive subdivision, or precede species formation without being its necessary cause. Consequently, neither phenotypic bimodality nor the temporal precedence of ecological differentiation is sufficient, by itself, to identify ecological divergence as the process that caused speciation.

Although ecological phenotype and reproductive compatibility are encoded by separate sets of loci, their evolutionary dynamics remain indirectly coupled through differential reproductive success and recombination. Environmental selection determines which ecological and reproductive-genetic backgrounds contribute to subsequent generations, whereas the compatibility threshold determines which pairs of individuals can reproduce and, consequently, which ecological chromosomes can be recombined. This architecture isolates an indirect contribution of environmental selection to reproductive isolation without invoking phenotype-based assortative mating, reinforcement, or magic traits \citep{servedio2011magic, nosil2012ecological}. This differs from models such as \cite{Dieckmann_2003}, in which ecological traits directly affect competitive interactions and assortative mating, thereby coupling ecological differentiation directly to reproductive isolation.

The homogeneous landscape provides the clearest demonstration that reproductive isolation can emerge independently of divergent environmental selection. Because all individuals experienced the same environment and were selected toward a common ecological optimum, diversification cannot be attributed to adaptation to contrasting niches. Nevertheless, restrictive genetic compatibility combined with local reproduction and limited dispersal generated multiple reproductively isolated species under both strong and weak selection. This result is consistent with theoretical studies showing that gradual spatial differentiation can produce reproductive subdivision in continuously distributed populations \citep{white1968models, gavrilets1998rapid, de2009global, martins2013, de2017barriers}. No species formed under permissive mating, which suggests that spatial structure and local reproduction disrupted reproductive connectivity only when the compatibility threshold was sufficiently restrictive.

Selection nevertheless modified the timing, phenotypic organization, and demographic consequences of spatial-genetic diversification. Strong selection rapidly concentrated phenotypes around the optimum and accelerated the increase in species richness, whereas under weak selection reproductive subdivision began while substantial phenotypic variation persisted inside the same environment. Strong selection also imposed a pronounced initial demographic cost. As phenotypes became concentrated around the optimum, mean fitness increased, reproductive failure declined, and compensatory recruitment allowed population size to recover. This recovery occurred substantially faster under restrictive than under permissive mating and coincided with a more rapid reduction in population-wide phenotypic variance. Selection therefore shaped how diversification unfolded even though the emergence of reproductive isolation depended primarily on restrictive compatibility, local reproduction, and limited dispersal.

These results connect the homogeneous regime to nonecological speciation and nonadaptive radiation, in which reproductively distinct lineages emerge without strong divergence in resource use or adaptation to contrasting niches \citep{czekanski2019ecology}. Secondary phenotypic peaks appeared in our homogeneous simulations only after reproductive subdivision and represented modest differences among species evolving around the same ecological optimum rather than adaptation to different environments. The resulting coexistence was therefore partially cryptic: reproductive differentiation was considerably stronger than ecological or phenotypic differentiation. Ring-species systems such as the \textit{Ensatina eschscholtzii} complex and the greenish warbler \textit{Phylloscopus trochiloides} provide a complementary, although not directly equivalent, empirical parallel. These systems illustrate how gradual differentiation across a geographically continuous distribution can coexist with reproductive discontinuities between some lineages \citep{moritz1992evolutionary, irwin2005speciation}. Together, these comparisons emphasize that continuous spatial variation need not imply continuous reproductive connectivity.

In the heterogeneous landscape, under the scenario of restrictive mating and strong selection (RS) and also under the scenario of permissive mating and weak selection (PW), the results were similar to those observed in the corresponding homogeneous scenarios. A different causal relationship emerged under permissive mating and strong selection. In the PS scenario, two environment-associated phenotypic modes developed while the population remained a single reproductive species for approximately $1,400$ generations. The permissive compatibility threshold continued to connect individuals across the environmental boundary, allowing recombination and the repeated production of intermediate phenotypes. Strong selection reduced the persistence and reproductive contribution of these intermediates until the reproductive-compatibility network separated into two disconnected components. Under weak selection and permissive mating, intermediate phenotypes experienced only small fitness penalties, reproductive-genetic connectivity persisted, and no species formed. Among the regimes examined, PS therefore most closely represents ecological speciation because reproductive isolation under permissive mating emerged only when divergent selection was strong \citep{Dieckmann_1999, nosil2012ecological}.

This transition did not require ecological phenotype to influence mate choice or compatibility directly. Instead, selection altered the demographic and genetic contributions of individuals dispersing, surviving, and reproducing near the environmental boundary. By reducing the contribution of maladapted intermediate phenotypes, ecological filtering weakened effective reproductive exchange between environments and allowed differences at the reproductive chromosome to accumulate until the compatibility network could no longer bridge individuals on the different environments. The prolonged transient before species formation reflects the continued opposition of dispersal, recombination, and permissive mating to this filtering process.

Ecology-first pathways in natural populations often include mechanisms that connect adaptation to reproduction more directly than in our model. In the apple maggot fly \textit{Rhagoletis pomonella}, host adaptation is associated with habitat-related mating and temporal isolation \citep{feder1994host, linn2004postzygotic}, whereas in \textit{Howea belmoreana} and \textit{H. forsteriana}, adaptation to contrasting soils is associated with differences in flowering time and reduced offspring fitness \citep{savolainen2006sympatric, hipperson2016ecological}. Lake whitefish similarly combine ecological and genomic differentiation with multiple reproductive barriers, ongoing gene flow, and, in some populations, secondary contact \citep{bernatchez2010origin}. By excluding components such as habitat preference, phenotype-based assortative mating, and reinforcement, the PS scenario isolates what selection against maladapted phenotypes can accomplish through ecological filtering alone. These other mechanisms should generally strengthen the association between ecological differentiation and reproductive isolation ~\citep{servedio2011magic, smadja2011framework} and may shorten the transition observed under PS. In the lake whitefish case ecological and nonecological stages of divergence are combined. Divergence among populations in separate glacial refugia preceded secondary contact, after which ecological differentiation, character displacement, and additional reproductive barriers evolved in sympatry \citep{bernatchez2010origin}. Present-day associations among habitat use, ecological phenotype, genomic differentiation, and reproductive isolation therefore reflect both earlier geographical divergence and subsequent ecological evolution. Such systems reinforce the central inference from our simulations: contemporary ecological differentiation need not identify the process that initiated lineage separation.

In contrast to PS, the RW scenario revealed the opposite temporal sequence in the heterogeneous landscape. Under restrictive mating and weak selection, reproductive subdivision began while the population-wide phenotypic distribution remained broad and only weakly associated with the two environments. Once reproductive components had formed, their ecological chromosomes could evolve more independently, and species-specific phenotypes gradually shifted toward the corresponding local optima. Environment-associated bimodality therefore emerged after reproductive subdivision had begun rather than initiating it. This sequence is consistent with evolutionary histories in which nonecological divergence is followed by ecological differentiation or character displacement \citep{czekanski2019ecology}. The comparison between RW and PW is particularly informative: permissive mating maintained reproductive exchange and a single intermediate phenotypic distribution, whereas restrictive mating allowed reproductively disconnected lineages to respond separately to the two environments. Similar long-term bimodality could therefore conceal fundamentally different historical relationships between ecological differentiation and reproductive isolation.

Phenotypic evolution under RW also failed to converge to a stationary pair of peaks at the environmental optima. Species-specific distributions continued to shift in position, width, and relative abundance for at least $10{,}000$ generations, even though reproductive isolation and total population size remained stable. Weak selection provided only a limited restoring force toward the local optima, allowing mutation, genetic drift, local reproduction, and spatial structure to continually reorganize ecological variation within reproductively disconnected lineages. Temporal phenotypic changes in natural populations are commonly interpreted as responses to fluctuating selection, whose strength and direction vary through time \citep{siepielski2009s}. Seasonal reversals in allele-frequency change in \textit{Drosophila melanogaster}, for example, have been associated with temporally varying environmental conditions and selection \citep{bitter2024continuously}. Our results reveal a distinct mechanism: persistent phenotypic fluctuations can arise even when the environmental optima and selection strength remain constant. Temporal variation in phenotype therefore does not necessarily imply temporal variation in selection. A single temporal sample may also capture species-specific phenotypic modes displaced from their local optima, obscuring both the constancy of the environment and the persistence of reproductive isolation.  A related caution is provided by the \textit{Desmognathus} complex, where present-day phenotypic differentiation does not map straightforwardly onto lineage history and may have arisen after lineage divergence \citep{jones2018genomic}. A testable prediction of this mechanism is that, under relatively weak divergent selection, repeated temporal sampling should reveal shifts in species-specific phenotypic means even when environmental conditions remain stable and reproductive isolation persists.

Future extensions should examine whether the causal distinctions identified here persist when ecological and reproductive loci are physically linked, ecological traits affect mating or habitat preference, environmental optima fluctuate or move through space, and reproductive isolation is evaluated using alternative operational criteria.

\section*{Acknowledgments}
%\begin{acknowledgments}
%Acknowledgments were omitted due to the blind review process.
 We thank the Coaraci Supercomputer for computer time (Fapesp grant 2019/17874-0) and the Center for Computing in Engineering and Science at Unicamp (Fapesp grant 2013/08293). We thank the reviewers and the editor who challenged us during the review process, resulting in an improved text.
%\end{acknowledgments}

\section*{Funding}
This work was partly supported by FAPESP, grant  2021/14335-0 (MAMA), and CNPq, grants 420803/2025-4 (FMDM), 303814/2023-3 (MAMA) and 141590/2024-6 (LDH). 

\section*{Availability of data and materials}

Data and code supporting the results of this study are available from the Dryad Digital Repository: \noindent{\url{https://doi.org/10.5061/dryad.s1rn8pkqm}} and \noindent{\url{https://github.com/LaraHissa/Spenvironments.git}}

\section*{Conflict of interests}
The authors declare no conflict of interests.

\section*{Author contributions}

Lara Domingos Hissa: Conceptualization, Data curation, Investigation, Methodology, Project administration, Software, Writing - original draft, Writing - review and editing.\\

\noindent Marcus A.M. de Aguiar: Conceptualization, Project administration, Supervision, Writing - original draft, Writing - review and editing. \\

\noindent Flavia M. D. Marquitti: Conceptualization, Project administration, Supervision, Writing - original draft, Writing - review and editing. 

\newpage
%\bibliographystyle{apalike}
%\bibliography{refs}

\begin{thebibliography}{}
\bibitem[White, 1968]{white1968models}
White, M. (1968).
\newblock Models of speciation: New concepts suggest that the classical
  sympatric and allopatric models are not the only alternatives.
\newblock {\em Science}, 159(3819):1065--1070.
\end{thebibliography}

\newpage

\includepdf[pages=-]{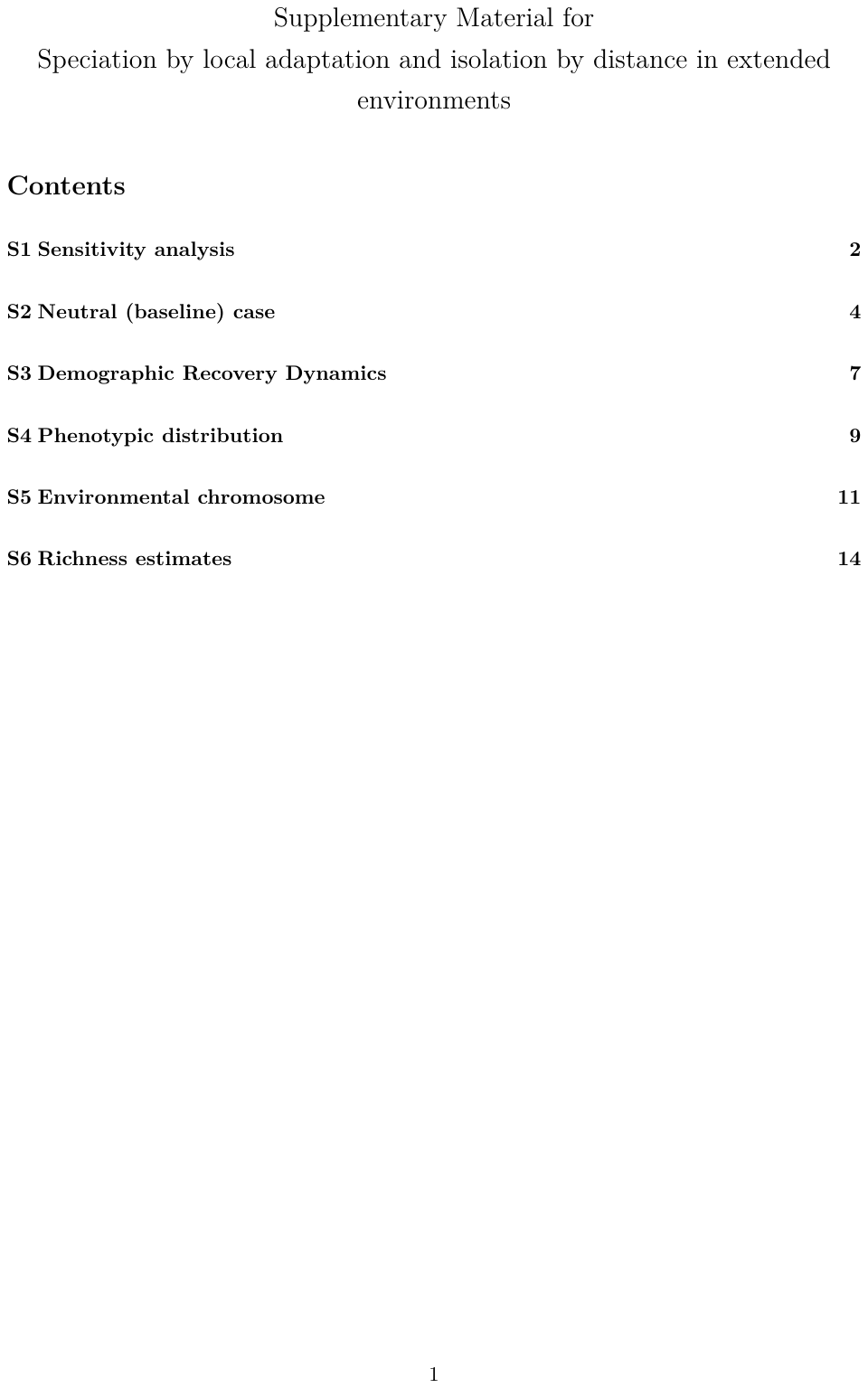}

%\AddToHook{enddocument/afteraux}{%
%\immediate\write18{
%cp output.aux speciation_rev2.aux
%}%
%}

 \end{document}